\documentclass[nofootinbib,showpacs,aps,pra,floatfix,article,preprint,longbibliography]{revtex4-2}

\usepackage[english,french]{babel}
\usepackage[utf8]{inputenc}
\usepackage{amsmath,amssymb,amsfonts,physics}
\usepackage{ragged2e}
\usepackage{graphicx}
\usepackage [T1] {fontenc}
\usepackage{bbold}
\usepackage{cases}
\usepackage{subcaption}
\usepackage[all]{xy}
\usepackage{hyperref}
\usepackage[usenames,dvipsnames]{xcolor}
\usepackage{tabularx}
\usepackage{cleveref}
\usepackage{dutchcal}
\usepackage[normalem]{ulem}
\usepackage{url}
\urldef\myurl\url{https://twitter.com/AmichaiStein1/status/1140374111258140673?ref_src=twsrc%5Etfw%7Ctwcamp%5Etweetembed%7Ctwterm%5E1140374111258140673%7Ctwgr%5E%7Ctwcon%5Es1_&ref_url=https%3A%2F%2Fwww.indiatimes.com%2Ftrending%2Fhuman-interest%2Fcrowd-of-protesters-in-hong-kong-parts-like-the-red-sea-to-make-way-for-ambulance-wins-hearts-369362.html}

\hypersetup{
    colorlinks=true,
    linkcolor=blue,
    filecolor=magenta,      
    urlcolor=blue,
    pdftitle={Overleaf Example},
    pdfpagemode=FullScreen,
    }
\urlstyle{same}

\newcommand{\orcid}[1]{\href{https://orcid.org/#1}{\includegraphics[width=10pt]{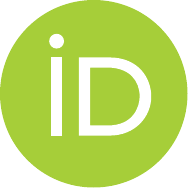}}}

\usepackage{hyperref}
\usepackage{amsfonts, vmargin}
\usepackage[us]{datetime}

\newcommand{\bx}{{\bf x}}
\newcommand{\bv}{{\bf v}}

\newcommand{\Phier}{\Phi_{\rm er}}
\newcommand{\Ger}{\Gamma_{\rm er}}

\newcommand{\m}{{m}}
\newcommand{\hl}{\xi} 
\newcommand{\sv}{c_s} 
\newcommand{\ba}{{\bf a}} 

\begin{document}
\begin{flushleft}
{\Large
\textbf\newline{\bf Pedestrians in static crowds are not grains, but game players}
}
\newline
\bigskip
\\
Thibault Bonnemain\orcid{0000-0003-0969-2413}\textsuperscript{1*},
Matteo Butano\orcid{0000-0002-8114-9725}\textsuperscript{2},
Théophile Bonnet\textsuperscript{3,2,\textcurrency},
I\~naki Echeverr\'ia-Huarte\orcid{0000-0001-6992-0815}\textsuperscript{4},
Antoine Seguin\textsuperscript{5},
Alexandre Nicolas\orcid{0000-0002-8953-3924}\textsuperscript{6},
Cécile Appert-Rolland\orcid{0000-0002-0985-6230}\textsuperscript{3},
Denis Ullmo\orcid{0000-0003-1488-0953}\textsuperscript{2}
\\
\bigskip
\textbf{1} Department of Mathematics, Physics and Electrical Engineering, Northumbria University, Newcastle upon Tyne, United Kingdom
\\
\textbf{2} Université Paris-Saclay, CNRS, LPTMS, 91405, Orsay, France
\\
\textbf{3} Université Paris-Saclay, CNRS, IJCLab, 91405, Orsay, France
\\
\textbf{4} Laboratorio de Medios Granulares, Departamento de Física y Matemática Aplicada, Univ. Navarra, 31080, Pamplona, Spain
\\
\textbf{5} Université Paris-Saclay, CNRS, FAST, 91405, Orsay, France
\\
\textbf{6} Institut Lumi\`ere Mati\`ere, CNRS \& Universit\'e Claude Bernard Lyon 1, 69622, Villeurbanne, France
\\
\bigskip
\textcurrency Current address: Université Paris-Saclay, CEA, Service d’Etudes des Réacteurs et de Mathématiques Appliquées, 91191, Gif-sur-Yvette, France

* Corresponding author: thibault.bonnemain@northumbria.ac.uk

\end{flushleft}

\section*{Abstract}
\justifying
The local navigation of pedestrians amid a crowd is generally believed to involve no anticipation beyond (at best) the avoidance of the most imminent collisions.
We show that current models rooted in this belief fail to reproduce
some key features experimentally evidenced  when a dense static crowd is crossed by an intruder. We identify the missing ingredient as the pedestrians' ability to plan their motion well beyond the next interaction, whence they may accept to move towards denser regions for a short time. To account for this effect, we introduce a minimal model based on mean-field game theory, which proves remarkably successful in replicating the  aforementioned observations as well as  other daily-life situations involving collective behaviour in dense crowds, 
such as partial metro boarding. This demonstrates the ability of 
game approaches to capture the anticipatory effects at play in operational crowd dynamics.

\vspace{2cm}

\newpage

\section{Introduction}
Although crowd disasters (such as the huge stampedes that grieved the Hajj in 1990, 2006 and 2015 \cite{helbing2007dynamics}) are
more eye-catching to the public, the dynamics of pedestrian crowds are also of great relevance in less dire circumstances. 
They are central when it comes to designing and dimensioning busy public facilities, from large transport hubs to entertainment venues, and  optimising the flows of people. Modelling pedestrian motion in these settings is a multi-scale endeavour, which requires determining where people are heading for (\emph{strategic} level),  what route they will take (\emph{tactical} level), and finally how they will move along that route in response to interactions with other people (\emph{operational} level)  \cite{hoogendoorn2004pedestrian}. 
The \emph{strategic} and \emph{tactical} levels typically involve some planning in order to make a choice among a discrete or continuous set of options, such as targeted activities, destinations~\cite{hoogendoorn2004pedestrian}, paths (possibly knowing their expected level of congestion) \cite{jiang2016comparison}, or, in the context of evacuations, egress alternatives~\cite{mesmer2014incorporation,jiang2016comparison}. These choices are often handled as processes of maximisation (minimisation) of a utility (cost), which may depend on lower-level information such as pedestrian density or streaming velocity~\cite{best2014,vangoethem2015}.

The \emph{operational} level deals with much shorter time scales and is generally believed to involve no planning ahead.
Anticipatory effects are thus merely neglected in so-called  \emph{reactive} models,
especially at high densities, possibly with the lingering idea that mechanical forces then prevail. For example, the popular social force model of Helbing and Molnar \cite{helbing1995social}, still at the heart of several commercial software products, combines contact forces and pseudo-forces (``social'' forces) which, in the original implementation, are only functions of the agents' current positions (and possibly orientations). Some degree of anticipation has since been introduced into these models to better describe collision avoidance, e.g., 
by making the pseudo-forces depend on future positions rather than current ones  \cite{zanlungo2011social,karamouzas2014universal}.
In a dual approach, the most imminent collisions can be avoided by scanning the whole velocity space \cite{vandenberg_l_m2008,paris2007pedestrian,karamouzas2017implicit} or a subset of it~\cite{moussaid2011simple} in search of the optimal velocity. In order to handle navigation through dense crowds, anticipated collisions beyond the most imminent one \cite{bruneau_p2017} or, at a more coarse-grained scale, local density inhomogeneities \cite{best2014} can be taken into
account in the optimisation. All these dynamic models, at best premised on a constant-velocity hypothesis, owe their high computational tractability to their relative shortsightedness: The simulated agents do not plan ahead in interaction with their counterparts.

In this paper, we argue that, even at the \emph{operational} level, crowds in some daily-life circumstances display signs of anticipation
that may elude the foregoing short-sighted models;
this will be exemplified by the recently studied response of a  {dense} static crowd when crossed by an `intruder' \cite{nicolas2019mechanical,kleinmeier2020agent}. We purport to show that a minimal game theoretical approach,
made tractable thanks to an elegant analogy between its mean-field formulation \cite{LasryLions2006-1,LasryLions2006-2,Huang2006}
and 
Schr\"odinger's equation \cite{Gueant2012,ULLMO20191}, can replicate the empirical observations for this example case, provided that it accounts for the anticipation of future costs. Beyond that particular example, the approach efficiently captures certain behaviours of crowds at the interface between the \emph{operational} and \emph{tactical} levels  { that are crucial to consider in attempts to improve the security of dense crowds}.

\section{Crossing a static crowd}

\begin{figure}
	\includegraphics[width=\textwidth]{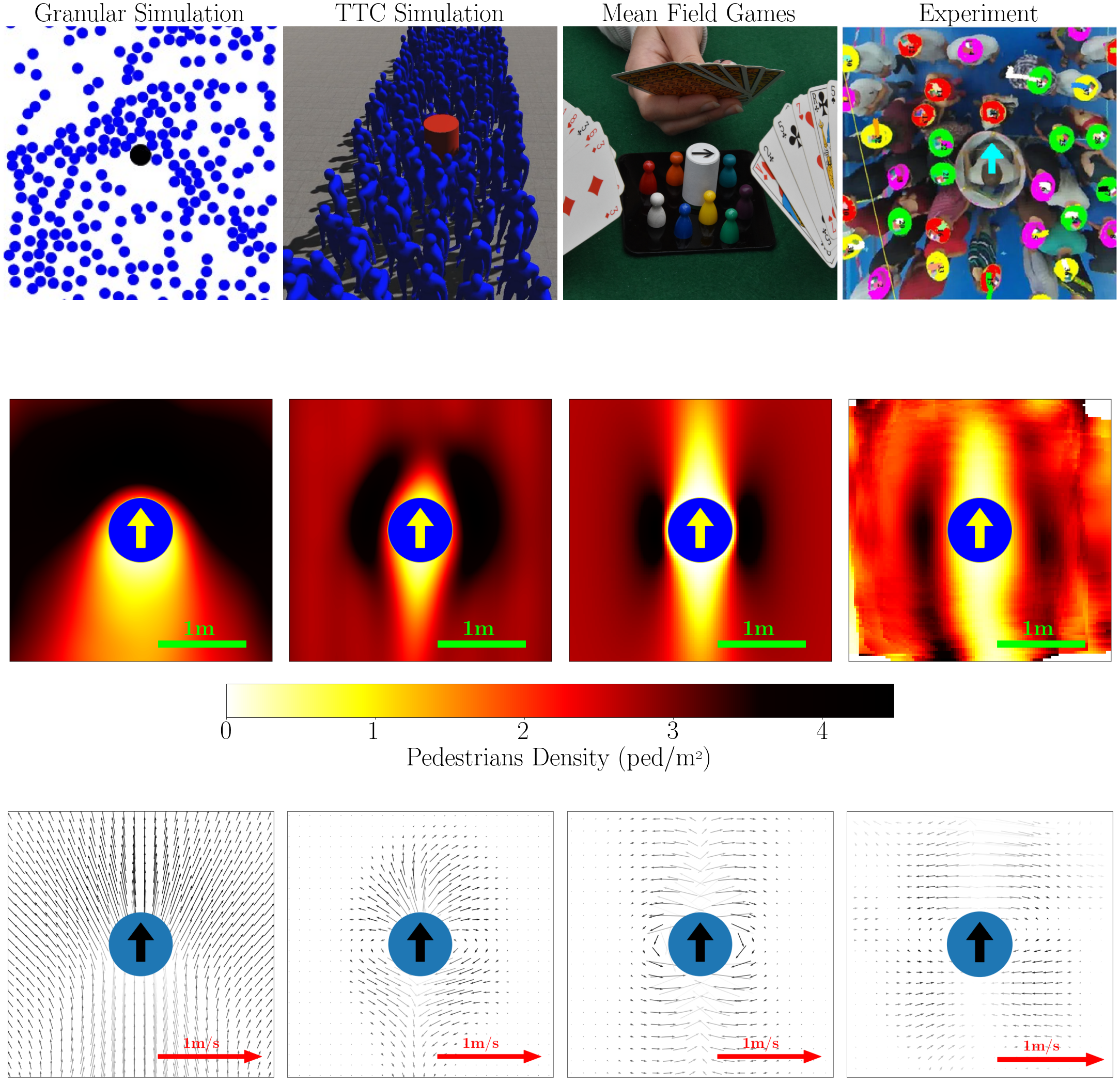}
\caption{
Density (middle row) and velocity (bottom row) fields induced in a static crowd by a cylindrical intruder that crosses it; the transparency of the velocity arrows is linearly related to the local density. (Column 1) Simulations of a mono-layer of vibrated disks. (Column 2) Simulations of an agent-based model
wherein agents may anticipate the most imminent collision. 
All fields have been averaged over many realisations. The snapshot illustrating the agent-based model was rendered using the Chaos visualisation software developed by INRIA (https://project.inria.fr/crowdscience/project/ocsr/chaos/). (Column 3) Results of the mean-field game introduced in this paper. (Column 4) Controlled experiments of  \protect{\cite{nicolas2019mechanical}}. Note the relatively symmetrical density dip in front and behind the intruder, as well as the transverse moves. (Columns 1-3) The crowd's density and intruder's size have been adjusted to match the experimental data (average density of 2.5 ped/m$^2$). Details of simulations and videos showcasing time evolution can be found in SI.}\label{fig:intro}
\end{figure}

Crossing a static crowd is a common experience in busy premises, from standing concerts and festivals to railway stations. Recently, small-scale controlled experiments \cite{nicolas2019mechanical,appert2020experimental} shed light on trends
that robustly emerged in the response of 
a crowd crossed by a cylindrical intruder, as
displayed in Fig.~\ref{fig:intro} (right column).
The induced response consists of a fairly symmetric density field around the intruder, displaying depleted zones both upstream and
downstream from the intruder, as well as higher-density regions on the sides. Indeed the crowd's displacements are mostly transverse: pedestrians tend to simply step aside.  Incidentally, a qualitatively similar response was filmed at much larger scale in a dense crowd of protesters in Hong-Kong, which split open to let an ambulance through~\cite{lien_yt}.

Such features strongly depart from the mechanical response observed \emph{e.g.}
in experiments \cite{seguin2011dense,seguin2013experimental} or simulations \cite{seguin2016clustering} of penetration into a 
granular mono-layer
below jamming, where grains  are pushed forward by the intruder (see Figs.~\ref{fig:intro} (left column)) and
   accumulate downstream, instead of moving crosswise. More worryingly, these ``mechanical'' features \cite{raj2021moving}
are also observed in simulations of pedestrian dynamics performed with
  the social-force model \cite{helbing1995social}, which rests on tangential and normal forces at contact and radial repulsive forces for longer-ranged interactions.
  
 Introducing collision anticipation in the pedestrian model helps
 reproduce the opening of an agent-free `tunnel' ahead of the intruder,
 as illustrated with a `time to (first) collision' model (second column of Fig.~\ref{fig:intro}) directly inspired from \cite{karamouzas2017implicit}, details of which can be found in SI. 
 However, even though the displacements need not align with the contact forces in this
 agent-based model,
 the displacement pattern diverges from the experimental observations, with
 streamwise (walk-away) moves that prevail over transverse (step-aside) ones. Indeed, 
such models rely on `short-sighted' agents, who do not see past the most imminent collision
expected  from  constant-velocity extrapolation. 

Results may vary with the specific collision-avoidance model and the selected parameters. Yet, our inability to reproduce
prominent experimental features suggests that an ingredient is missing in these approaches based on short-time (first-collision) anticipation.

\section{A game theoretical approach to account for low-level planning}

To bring in the missing piece, we start by noticing that the observed behaviours are actually most intuitive: Pedestrians anticipate that it will cost them less effort to step aside and then resume their positions, even if it entails enduring high densities for some time,
than to endlessly run away from an intruder that will not deviate from its course.
But accounting for this
requires a change of paradigm compared to the foregoing approaches.
Game theory is an adequate framework to handle the conflicting impulses of interacting agents endowed with planning capacities: agents are now able to 
optimise their strategy taking into account the choices (or strategies) of others. So far, its use in pedestrian dynamics has mostly been restricted to evacuation tactics in discrete models ~\cite{heliovaara2013patient,bouzat2014game,mesmer2014incorporation}. Unfortunately, the problem becomes intractable when the number of interacting agents grows.

To overcome this quandary, we turn to Mean Field Games (MFG), introduced
by Lasry and Lions \cite{LasryLions2006-1,LasryLions2006-2} as well as Huang et al. \cite{Huang2006} in the wake of the mean-field approximations of statistical mechanics,
and since used in a variety of fields, ranging from finance \cite{Lachapelle2014,Cardaliaguet2017,Carmona-ctrl2013} to economics \cite{Achdou2016,GueantLasryLions2010,Achdou2014}, epidemiology \cite{LaguzetTurinici2015,djidjou2020optimal,elie2020contact}, sociology \cite{LachapelleWolfram2011,achdou2017mean,LaguzetTurinici2015}, or engineering \cite{KizilkaleMalhame2016,kizilkale2019integral,WirelessNetwork}.
While applications of MFG to crowd dynamics 
have already been proposed
\cite{Gueant2015,lachapelle2011mean,jiang2016macroscopic,jiang2016comparison},   our goal here is to demonstrate the practical relevance of this  approach at the \emph{operational} level, using an elementary MFG belonging to one of the first class of models introduced by Lasry and Lions \cite{LasryLions2006-1}, and which can be thoroughly analysed thanks to its connection with the nonlinear Schrödinger equation.

In the mean field approximation, the “N-player” game is replaced by a generalized Nash equilibrium \cite{kreps1989nash} where indiscriminate microscopic agents play against a macroscopic state of the system (a density field) formed by the infinitely many remaining agents. Consider a large set of pedestrians, the agents of our game,  characterised by their spatial position (state variable) ${\bf X}^i\equiv(x^i,y^i)\in \mathbb{R}^2$, which we assume follows Langevin dynamics, viz.,
\begin{equation}\label{eq:lang}
d{\bf X}^i_t=\ba^i_tdt+\sigma d {\bf W}^i_t \; ,
\end{equation}
where the drift velocity (\textit{control} variable) $\ba^i_t$ reflects the agent's strategy.
In \eqref{eq:lang}, $\sigma$ is a constant and components of ${\bf W}^i$ are independent white noises of variance one accounting for unpredictable events.  All agents are supposed identical, apart from their initial positions ${\bf X}^i(t=0)$ and realisations of ${\bf W}^i$. 

Each agent strives to adapt their velocity $\ba^i_t$ in order
 to minimise a cost functional that we assume to take the simple form
\begin{equation}\label{eq:cost}
c[{\bf a}^i](t,{\bf x}^i_t) =    
\left \langle \displaystyle \int_{t}^T \left[ \frac{\mu \ba^2}{2} -
\left(g\m_t({\bf x}) + U_0({\bf x} \! - \! {\bf v} t )\right) \right] d\tau \right \rangle  \; ,
\end{equation}
where the average denoted by  $\langle \cdot \rangle$ is performed over all realizations of the noise for trajectories starting at ${\bf x}^i_t$ at time $t$. 
In this expression, the term ${\mu \ba^2}/{2}$, akin to a kinetic energy, represents the efforts required by the agent to enact their strategy (how much/how fast they have to move in this case), while the interactions with the other agents via the empirical density
$\m^{(e)}(t,{\bf x}) = \sum_i \delta ({\bf x} - {\bf X}^i(t))/N$
are controlled by a parameter $g<0$. Finally, the space occupied by the intruding cylinder, which moves at a velocity  ${\bf v}=(0,v)$, is characterised by
a `potential' $U_0({\bf x}) = V_0 \Theta(\norm{\bf x} - R  )$ that tends to  $V_0 \to -\infty$ inside the radius $R$ of the cylinder and is zero elsewhere. Agents need to balance those three terms \emph{over the whole duration $T$ of the game}, which enables them to make costly, but temporary  moves 
if they lower the overall cost.
For example, depending on the parameters, stepping aside into a high density region
(a cost-inefficient strategy \emph{a priori})
to let the intruder through may prove overall more efficient than running away from it; the first strategy implies paying a high cost upfront, but nothing afterwards, while the second implies paying a comparatively low cost
that however extends over the whole duration of the game, resulting in a potentially worse pay-off.

In the presence of many agents, the density self-averages to
$\m(t,{\bf x}) = \langle \m^{(e)}(t,{\bf x})\rangle_{\rm noise}$ and the optimization problem \eqref{eq:cost} does not feature explicit coupling between agents anymore. It can then be solved by introducing the value function $\displaystyle u(t,{\bf x})=\min_{a(\cdot)} c[\ba](t,{\bf x})$, which obeys a Hamilton-Jacobi-Bellman [HJB] equation \cite{bellman1957dynamic,LasryLions2006-2}, with an optimal control given by $\ba^*(t,{\bf x}) = - \nabla u(t,{\bf x}) / \mu$.  Consistency imposes that $\m(t,{\bf x})$ is solution of the Fokker-Planck [FP] equation associated with \eqref{eq:lang}, given the drift velocity $\ba(t,{\bf x}) = \ba^*(t,{\bf x})$. As such, MFG can be reduced to a system of two coupled partial differential equations \cite{Gueant2012,LasryLions2006-1,LasryLions2006-2,ULLMO20191}.
\begin{equation}
\label{MFG}
\left\{
\begin{aligned}
\partial_t u(t,{\bf x}) & = \frac{1}{2\mu}\left[\nabla u(t,{\bf
    x})\right]^2 - \frac{\sigma^2}{2}\Delta u(t,{\bf x}) + g\m(t,{\bf x}) +
U_0(\bx - {\bf v}t) \qquad \mbox{[HJB]}\\
\partial_t \m(t,{\bf x}) &= \frac{1}{\mu}\nabla\left[\m(t,{\bf x})\nabla
  u(t,{\bf x})\right] + \frac{\sigma^2}{2}\Delta \m(t,{\bf x})  \qquad \qquad \qquad
\qquad \qquad \mbox{[FP]}
\end{aligned}
\right. \; ,
\end{equation}

The atypical ``forward-backward'' structure of Eqs.~(\ref{MFG}), highlighted by the opposite signs of Laplacian terms in the two equations,
accounts for anticipation. The boundary conditions epitomise this structure: based on \eqref{eq:cost}, the value function has terminal condition $u(t=T,{\bf x})=0$, while the density of agents evolves from a fixed initial distribution $\m(t=0,{\bf x})=\m_0({\bf x})$. 
In previous work, we have evinced a formal, but insightful
mapping of these MFG equations onto a nonlinear Schr\"odinger equation (NLS) \cite{Gueant2012,ULLMO20191,bonnemain2019universal}, which has been
studied for decades in fields ranging from non-linear optics
\cite{Kaup1990} to Bose-Einstein condensation
\cite{pitaevskii2016bose} and fluid dynamics
\cite{NLSfluid}. 

We perform a change of variables $(u(t,\bx),\m(t,\bx)) \mapsto (\Phi(t,\bx), \Gamma(t,\bx))$ through $u(t,\bx) = - \mu\sigma^2 \log \Phi(t,\bx)$, $\m(t,\bx)  = \Gamma(t,\bx) \Phi(t,\bx)$ \cite{ULLMO20191}.
The first relation is the usual Cole-Hopf transform
\cite{ColeHopf}; the second corresponds to an  "Hermitization" of
Eqs.~(\ref{MFG}). In terms of the new variables
$(\Phi,\Gamma)$, the MFG equations read
\begin{equation} \label{eq:NLS}
\left\{
\begin{aligned} 
- \mu\sigma^2\partial_t\Phi   & =\frac{\mu\sigma^4}{2}\Delta\Phi + (U_0+g\Gamma\Phi)\Phi \\
+ \mu\sigma^2\partial_t\Gamma & =\frac{\mu\sigma^4}{2}\Delta\Gamma + (U_0+g\Gamma\Phi)\Gamma 
\end{aligned} \right. \; .
\end{equation}
Except for the missing imaginary factor associated with time derivation, these equations have exactly the structure of NLS describing the evolution of a quantum state $\Psi(t,{\bf x})$ of a Bose-Einstein condensate, with formal correspondence 
$\Psi \rightarrow \Gamma$,
$\Psi^* \rightarrow \Phi$ and
$\rho \equiv||\Psi||^2\rightarrow \m \equiv \Phi\Gamma$. 
This system, however, retains the forward-backward structure of MFG evidenced by mixed initial and final boundary conditions $\Phi(T, \bx)=1$, $\Gamma(0, \bx)\,\Phi(0,\bx)={\m_0(\bx)}$.
Several methods have been developed to deal with NLS and most can be 
leveraged to tackle the MFG problem \cite{ULLMO20191,bonnemain2019schrodinger}.

Self-consistent solutions of Eqs.~(\ref{eq:NLS}) are obtained by iteration over a backward-forward scheme. A video illustrating the evolution of the agents' density for a particular set of parameters, as well as details about the numerical scheme, can be found in SI.

Focusing on the \emph{permanent} regime (a.k.a. the {\em ergodic state} \cite{Cardaliaguet2013}) , rather than on the transients associated with the intruder's entry or exit,
further simplifies the resolution.
In this regime, defined by time-independent density and velocity fields in the intruder's frame, the auxiliary functions $\Phi$ and $\Gamma$ are not constant in time, but they assume the trivial dynamics $\Phi(t,\bx) = \exp[\lambda t/\mu \sigma^2] \Phi_{\rm er}$ and $\exp[-\lambda t/\mu \sigma^2] \Gamma_{\rm er}$ where, in the frame of the intruder, $\Phi_{\rm er}$ and $\Gamma_{\rm er}$ satisfy
\begin{equation}\label{eq:ergo}
  \left\{
\begin{aligned}
  &\frac{\mu\sigma^4}{2}\Delta{\Phier} - \mu\sigma^2 \bv \cdot\vec{\nabla}{\Phier} + [U_0(\bx) + g \m_{\rm er}] {\Phier} = -\lambda {\Phier} ,  \\
  &\frac{\mu\sigma^4}{2}\Delta{\Ger} + \mu\sigma^2 \bv \cdot\vec{\nabla}{\Ger} + [U_0(\bx) + g \m_{\rm er}] {\Ger} = - \lambda {\Ger} 
\end{aligned}
  \right. \; ,
\end{equation}
(with $\m_{\rm er} = \Phier \Ger$ independent of time).  Far from the intruder  $ U_0(\bx) = 0$, $m \simeq m_0$ and pedestrians  have constant velocity $-\bv$ in the intruder frame.  This imposes the asymptotic solutions $\Phier(\bx)$ $ =$ $\Ger(\bx)$ $ = \sqrt{m_0}$, from which $\lambda = -g \m_0 $.

\section{Results}

The ergodic Eqs.~(\ref{eq:ergo}) have two remarkable features:  (i) They give direct access to the permanent regime, and are straightforward to implement numerically since time dependence has disappeared. 
(ii) The solutions of Eqs.~(\ref{eq:ergo}) are entirely specified by two dimensionless parameters. 

Indeed, the intruder is characterised by its radius $R$ and its velocity $v$.
In the same way, pedestrians are characterized by a length scale $\hl=\sqrt{|\mu\sigma^4/2g m_0|}$, the distance over which the crowd density tends to recover its bulk value from a perturbation, a.k.a  \emph{healing length}, and a velocity scale $\sv=\sqrt{\left|{g\bar m_0}/{2\mu}\right|}$, the typical speed at which pedestrians tend to move\footnote{Note that $\mu \hl \sv = \mu \sigma^2$ has the dimension of an action and plays the role of $\hbar$ in the original nonlinear Schrödinger equation.}. Up to a scaling factor, solutions of Eqs.~(\ref{eq:ergo}) can be expressed as a function of the two ratios $\hl/R$ and $\sv/v$ instead  of depending on the full set of parameters $(R,v,\mu,\sigma,m_0,g)$, which facilitates the exploration of the parameter space.

\begin{figure}[ht]
	\includegraphics[width=\textwidth]{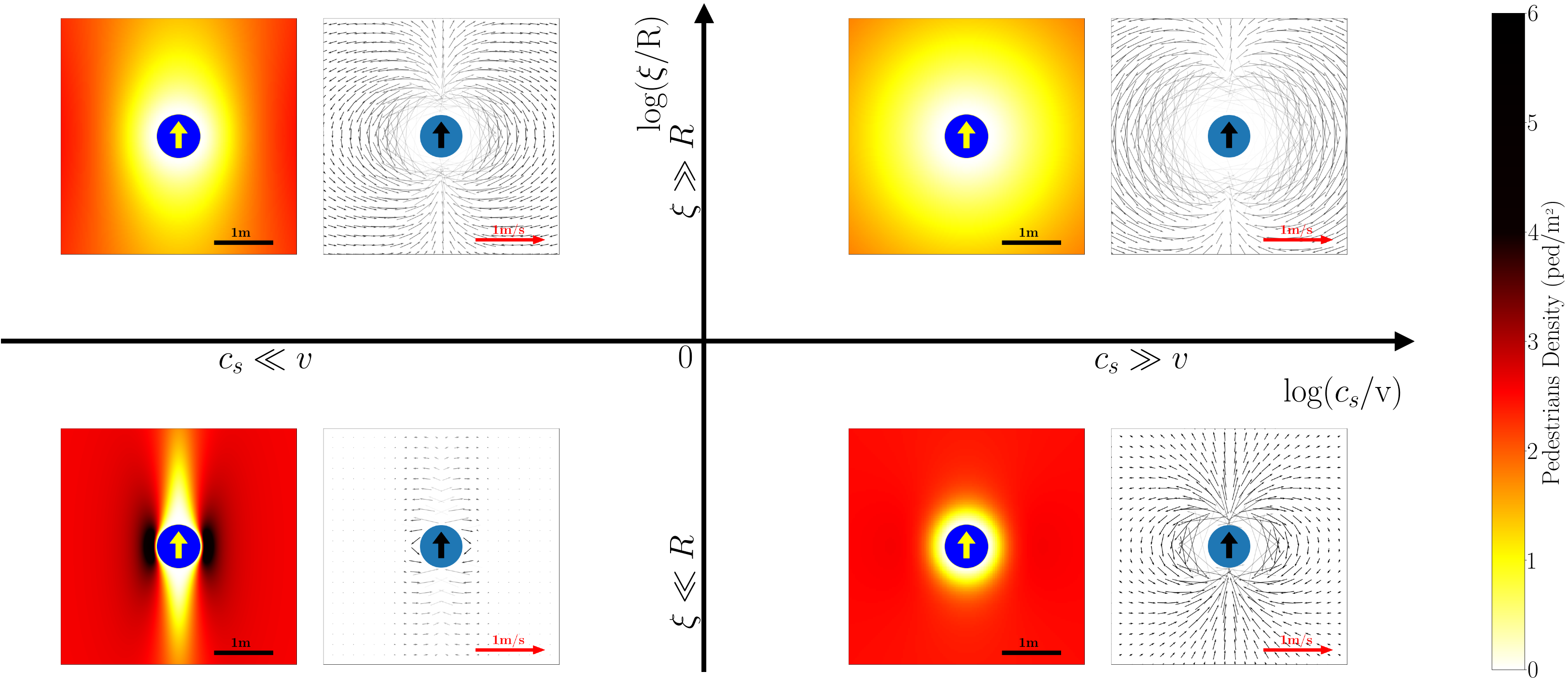}
    \caption{Typical density and velocity fields induced by the crossing intruder in the ergodic state, as predicted by the MFG model in 
    different regions of the parameter space. Parameters taken in the small $\sv/v$ and small $\hl/R$ quadrant display good visual agreement with the experimental data.}
    \label{fig:MFGzeta}
\end{figure}

Figure~\ref{fig:MFGzeta} presents typical density and velocity fields simulated in the ergodic state, for parameters selected in each quadrant of the reduced space parametrised by
$\log{\sv/v}$ and $\log{\hl/R}$ on the horizontal and vertical axis respectively.
Intuitively, one understands that $\sv$ governs the cost of motion for the agents
while $\hl$ gives the extent of the perturbation caused by the presence of the intruder. The main visual difference between the small and large $\sv/v$ cases is the change in rotational symmetry, a fact that reflects a more fundamental change in strategy. For large values of $\sv/v$ pedestrians do not mind  moving, and they rather try to avoid congested areas for as long as possible, thus creating circulation around the intruder, as shown in the velocity plots. On the other hand, for small values of $\sv/v$, moving fast costs more; therefore, in order to avoid the intruder, pedestrians
have to move earlier, and 
will accept to temporarily side-step into a more crowded area, thereby causing a stretch of the density along the vertical direction.
The experimental observations of \cite{nicolas2019mechanical} are best reproduced for small $\sv/v$ and small $\hl/R$ ($\sv = 0.11$ and $\hl = 0.15$), as shown in the third column of Fig.~\ref{fig:intro}.  Considering the minimalism of our MFG model, the obtained agreement is especially satisfying.

\section{Discussion}

The data plotted in Fig.~\ref{fig:intro} (third column) demonstrate that even basic MFG models can naturally capture and semi-quantitatively reproduce prominent features of the response of static crowds \cite{nicolas2019mechanical}, 
which may be out of reach of more short-sighted pedestrian dynamics models.

\begin{figure}[ht]
	\includegraphics[width=.34\textwidth]{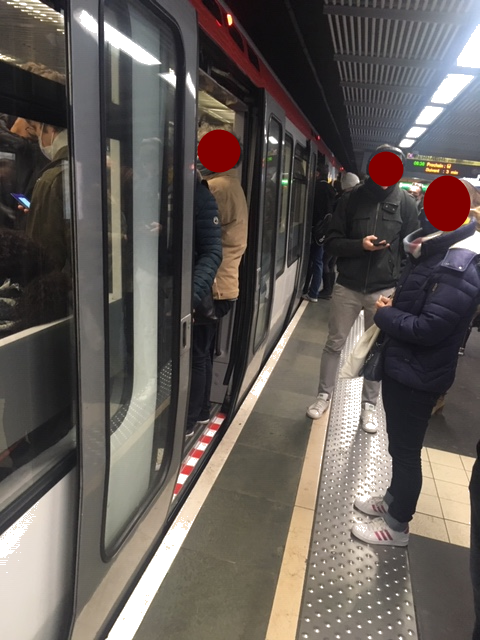}  \hspace{.5cm}
		\includegraphics[width=.53\textwidth]{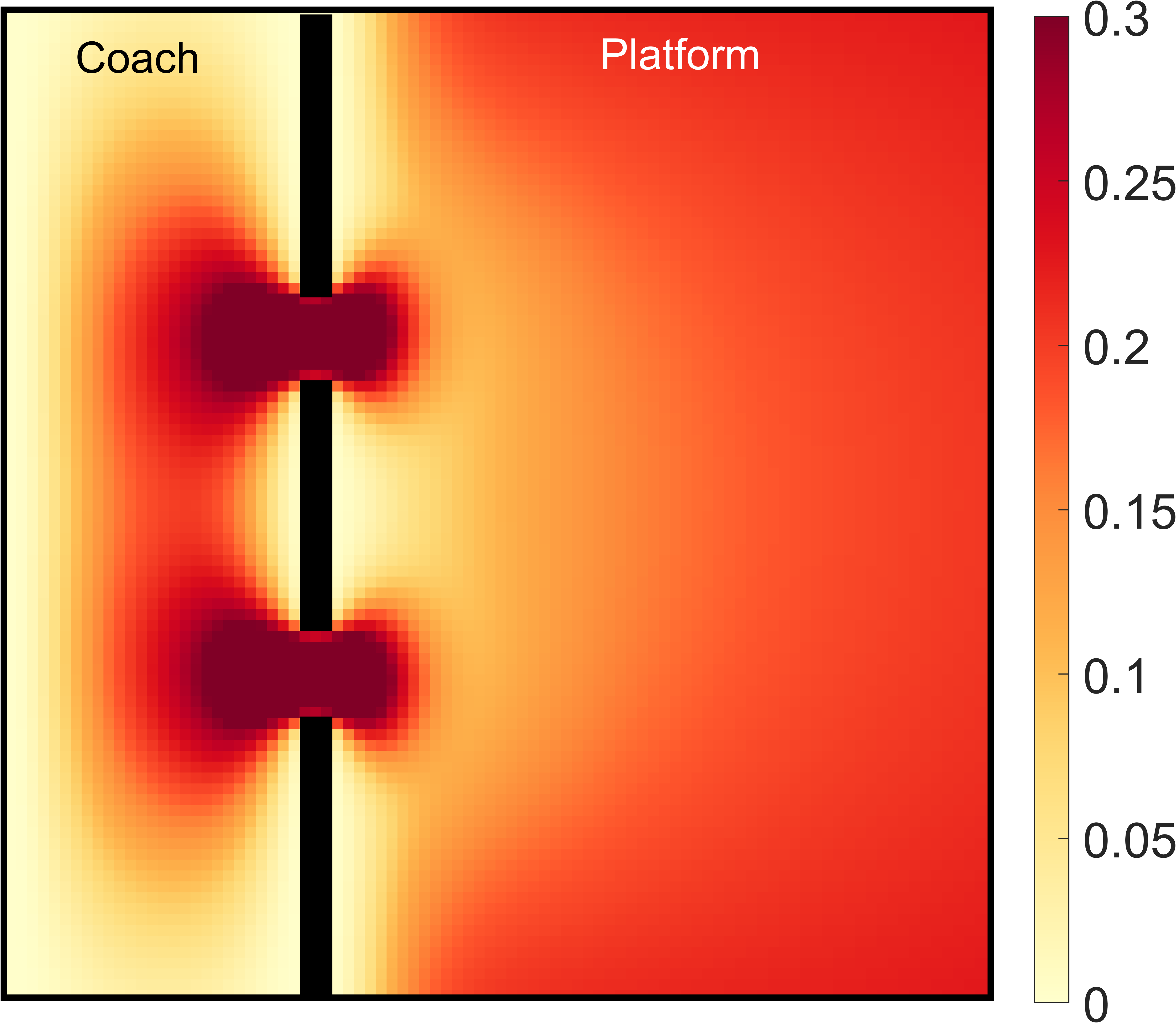} 
\caption{
Boarding a crowded metro coach at rush hour. Left:  Morning rush hour of November 18, 2021, on the platform of Metro A in Lyon, France. The doors are about to close and the gap between boarding passengers and those who preferred to wait for the next metro is clearly visible.
Right: Snapshot from a MFG simulation at $t=0.9T$. Players start uniformly distributed on the platform and would like to get on the coach before the doors close, at $t=T$. Just before that moment, the players closest to the doors choose to rush towards the coach and cram themselves in it despite the high density. Others prefer to stay on the platform (see SI for a movie of the whole process).}
\label{fig:metro}
\end{figure}

Beyond this particular example, MFG are also applicable to a broader array of crowd-related problems. This will now be illustrated by exploring the 
 daily-life situation of people waiting to board the coach in an underground station. This is readily achieved by suitably modifying the external potential $U_0(\bx)$ and the geometry of the system,
as shown on  Fig.~\ref{fig:metro}, and introducing a terminal cost $c_T(\bx)$ \cite{ULLMO20191,bonnemain2019schrodinger} that is lower
aboard the metro than on the platform. By solving the time dependent equations \eqref{eq:NLS}, we manage to reproduce the boarding process in a qualitatively realistic way,  up to the decision made
by some agents to stay on the platform rather than board the overcrowded metro. We believe this last point to be particularly interesting since this ``passive'' behaviour emerges naturally from our (anticipatory) game theoretical model, something that would be essentially impossible to implement without an ad hoc treatment in traditional approaches of crowd dynamics at the tactical level.

To conclude, let us recall that the foregoing results have been obtained with a simple, generic MFG model which depends linearly on density via $g \m(t,\bx)$.
This approximation can be refined and the MFG formalism is flexible enough to incorporate further elements to make it truer to life, including time-discounting effects \cite{frederick2002time,gomes2015economic} and congestion \cite{Dogbe2010,Gueant2015,Achdou2018}. Higher quantitative accuracy will be within reach of these more sophisticated approaches, possibly at the expense
of less transparent outcomes compared to the elementary model used here. For sure, MFG will struggle to capture a variety of problems of crowd
dynamics at the operational level, notably those for which the granularity of the crowd is central.
However, the fact that {\em even the simplest} of the Mean Field Game models is able to capture qualitative features that are missed not only by ``off-the-shelf'' commercial software, but also by a state-of-the-art `time to (first) collision' model including some anticipation, bolster the claim that  {\em optimization} and {\em anticipation} are essential ingredients for the description of  crowd dynamics  at the \emph{operational} level, and justifies to claim entry for Mean Field Game based approaches into the toolkit of practitioners of the field.

\section*{Methods}
\subsection*{Simulations}

The granular response (first column of Fig.~1) to the penetration of an intruder of diameter $D=2d=0.74$~m was obtained by simulating 
the dynamics of a two-dimensional layer of identical frictionless grains of diameter $d=0.37$~m with molecular dynamics. The interactions between grains were given by Hertzian contact forces $\displaystyle F_{ij}=k\zeta^{3/2}-\lambda \frac{d\zeta}{dt}$, where $\zeta$ is the interpenetration of the grains,  $k$ is the stiffness of the contact, and $\lambda$ is a damping coefficient. 

The agent-based simulation (second column of Fig.~1) is performed with a model based on anticipated times to collision (TTC), inspired by \cite{karamouzas2017implicit}. In this model, at each time step every agent $i$ selects their desired velocity $\boldsymbol{v'_p}$ as the minimum of an individual cost function (or pseudo-energy) $E_i[\boldsymbol{v'_p}]$; velocities $\boldsymbol{v'_p}$ that lead to a collision with another agent $j$ within a very short time horizon $\tau_{ij}$ (if $j$ keeps their current velocity) are penalized by 
a cost $E_i^{\text{TTC}}$ (reproduced from \cite{karamouzas2014universal}) in $E_i$ which becomes very large when one of the TTC $\tau_{ij}$ gets small. In addition to this TTC term,
the total cost $E_i$ includes (i) a driving term $E^{\text{target}}$, which assesses whether $\boldsymbol{v'_p}$ brings the agent closer to the destination, (ii) a term constraining the agent's speed, $E^{\text{speed}} \propto v'_p(\boldsymbol{v'_p}-\boldsymbol{v_p}^{\text{pref}})^2$, where $\boldsymbol{v_p}^{\text{pref}}$ is a comfortable walking speed, (iii) a term penalizing sudden changes in velocity, (iv) a repulsion term, $E^{\text{core-repulsion}}$ that is activated as soon as another agent steps into the private sphere of agent $i$ and then grows as the inverse of their mutual distance $f$.

MFG simulations are realised by numerically solving either Eqs.~(\ref{eq:NLS}) or (\ref{eq:ergo}). 

Further details about the different algorithms can be found in the SI.

\subsection*{Smoothing of the density and velocity fields}
The smooth velocity fields $\tilde{\boldsymbol{v}}$ shown on Fig.~1 were obtained by convoluting the discrete instantaneous experimental or numerical fields $\boldsymbol{v}(\boldsymbol{r})=\sum_i \boldsymbol{v}_i \delta(\boldsymbol{r}-\boldsymbol{r}_i)$ (where the sum runs over all particles $i$ and $\delta$ denotes a Dirac distribution) with a Gaussian kernel $\phi(\boldsymbol{r})=\frac{1}{2\pi r_c^2} \cdot \exp(-\frac{1}{2}\frac{\boldsymbol{r}^2} {r_c^2})$ with $r_c \approx 20 \mathrm{cm}$, viz.,
$\tilde{\boldsymbol{v}}(\boldsymbol{r})= \int d^2\boldsymbol{r'} \phi(\boldsymbol{r}-\boldsymbol{r'}) \boldsymbol{v}(\boldsymbol{r'})$. (In practice, the Gaussian kernel was truncated at $3 r_c$). The coordinates $\boldsymbol{r}$ were then re-centered around the intruder's position at each time frame and the
resulting fields were averaged over time. A similar smoothing process was used for the density fields.

\section*{Data Availability}
All study data are included in this article or the SI. Movies S1–S4 have been deposited in the Open Science Framework
(OSF) (Movie S1, \url{https://osf.io/cgs7y/}; Movie S2, \url{https://osf.io/te64f/}; Movie S3, \url{https://osf.io/vjzby/}; Movie S4, \url{https://osf.io/b7ep8/}).

\bibliographystyle{ieeetr}

\section*{Acknowledgments}

We acknowledge financial support for the internship of Theophile Bonnet by the ``Investissements d’Avenir'' of LabEx PALM (ANR-10-LABX-0039-PALM), in the frame of the PERCEFOULE project, as well as funding from the Hubert Curien Partnership France-Malaysia Hibiscus (PHC- Hibiscus) programme [203.PKOMP.6782005].

\section*{Author contributions}

T. Bonnemain, M. Butano and D. Ullmo provided expertise in MFG; I. Echeverr\'ia-Huarte, A. Nicolas, and C. Appert-Rolland provided expertise in crowd and pedestrian dynamics; A. Seguin provided expertise in granular materials; T. Bonnemain and T. Bonnet worked on the time-dependent MFG simulations; M. Buttano worked on the ergodic state MFG simulations; D. Ullmo and C. Appert-Rolland supervised trainees on MFG and provided financial support; A. Seguin worked on the granular material simulations; I. Echeverr\'ia-Huarte and A. Nicolas worked on TTC simulations; All contributors participated to the discussion regarding the physics of crowd dynamics and to the writing of the paper.

\section*{Competing interests}
The authors declare no competing interests.

\end{document}